\documentclass[11pt,preprint]{aastex}

\usepackage{natbib}

\newcommand{\sdssa}{SDSS J2357$-$0052}
\newcommand{\sdssb}{SDSS J1703$+$2836}
\newcommand{\kms}{~km~s$^{-1}$} 
\newcommand{\teff}{$T_{\rm eff}$}
\newcommand{\logg}{$\log g$}

\newcommand{\vt}{$v_{\rm micro}$}

\newcommand{\vh}{$V_{\rm helio}$}

\slugcomment{Submitted to ApJ Letters}

\shorttitle{R-process Enhancements in an Extremely Metal-Poor Subdwarf}
\shortauthors{Aoki et al.}

\begin{document}


\title{Extreme Enhancements of r-process Elements
in the Cool Metal-Poor Main-Sequence Star SDSS J2357$-$0052}


\author{Wako Aoki\altaffilmark{1,2}}
\affil{National Astronomical Observatory, Mitaka, Tokyo,
181-8588 Japan}
\email{aoki.wako@nao.ac.jp}
\altaffiltext{2}{Department of Astronomical Science, The Graduate
  University of Advanced Studies, Mitaka, Tokyo, 181-8588 Japan}

\author{Timothy C. Beers\altaffilmark{3}}
\affil{Department of Physics and Astronomy and JINA: Joint Institute for Nuclear
  Astrophysics, Michigan State University, East Lansing, MI
  48824-1116}
\email{beers@pa.msu.edu}

\author{Satoshi Honda\altaffilmark{4}}
\affil{Kwasan Observatory, Kyoto University, Ohmine-cho Kita Kazan,
  Yamashina-ku, Kyoto, 607-847, Japan; honda@kwasan.kyoto-u.ac.jp}

\and

\author{Daniela Carollo\altaffilmark{5,6}}
\affil{Research School of Astronomy \& Astrophysics, Australian National University\\ 
\& Mount Stromlo Observatory, Cotter Road, Weston, ACT, 2611, Australia}
\email{carollo@mso.anu.edu.au}
\altaffiltext{6} {INAF-Osservatorio Astronomico di Torino, Italy}




\begin{abstract}

We report the discovery of a cool metal-poor, main-sequence star exhibiting
large excesses of r-process elements. This star is one of two newly discovered
cool subdwarfs (effective temperatures of 5000~K) with extremely low metallicity
([Fe/H]$<-3$) identified from follow-up high-resolution spectroscopy of
metal-poor candidates from the Sloan Digital Sky Survey. {\sdssa} has
[Fe/H]$=-3.4$ and [Eu/Fe]$=+1.9$, and exhibits a scaled solar r-process
abundance pattern of heavy neutron-capture elements. This is the first example
of an extremely metal-poor, main-sequence star showing large excesses of
r-process elements; all previous examples of the large r-process-enhancement
phenomena have been associated with metal-poor giants. The metallicity of this
object is the lowest, and the excess of Eu ([Eu/Fe]) is the highest, among the
r-process-enhanced stars found so far. We consider possible scenarios to account
for the detection of such a star, and discuss techniques to enable searches for
similar stars in the future.

\end{abstract}





\section{Introduction}

Over the course of the past decade, a great deal of observational effort has
been dedicated to studies of the chemical enrichment of the early universe
through high-resolution spectroscopic measurements of the elemental abundances
of metal-poor stars in the Galaxy. Of particular interest are the extremely
metal-poor (EMP) stars, with [Fe/H]$\lesssim -3$~\footnote{[A/B] = $\log(N_{\rm
A}/N_{\rm B}) -\log(N_{\rm A}/N_{\rm B})_{\odot}$, and $\log\epsilon_{\rm A}
=\log(N_{\rm A}/N_{\rm H})+12$ for elements A and B.}, as such stars are
expected to have recorded the nucleosynthesis products associated with the first
generations of stars, perhaps even by individual supernovae, according to models
such as the supernova-induced, low-mass star formation scenario
\citep[e.g.][]{audouze95,shigeyama98}.

A number of interesting abundance patterns have been noted to occur for
individual EMP stars; the most remarkable example is perhaps the occurrence of
the so-called r-process-enhanced stars \citep[e.g. ][]{sneden03, cayrel01}. Some
EMP stars have been shown to exhibit over-abundances of the heavy
neutron-capture elements, such as Eu with respect to Fe, by more than an
order of magnitude, while the abundance patterns of their heavy elements agree
well with that of a scaled solar r-process component \citep{sneden08}. Such
stars may well represent the yields of an essentially pure r-process production
event, the yields of which were apparently not fully mixed into the interstellar
medium during early stages of the Galaxy's evolution.

The astrophysical site(s) of the r-process is/are not yet established with
confidence. However, an important clue may be the fact that all extremely
r-process-enhanced stars (r-II stars; Beers \& Christlieb 2005) identified to
date have been found in a relatively narrow metallicity range
($-3<$[Fe/H]$<-2.5$). The upper metallicity bound for the presence of r-II stars
is well constrained by measurements for a large sample of stars in this
metallicity range ([Fe/H]$\gtrsim -2.5$), and could be explained by the increase
of the Fe abundance with a limited supply of Eu from individual
supernovae\footnote{An exception is COS~82 in the dwarf spheroidal galaxy Ursa
Minor, which exhibits [Fe/H]$=-1.5$ and [r/Fe]$=+1.5$
\citep{aoki07}. However, such an object has not yet been found among the Milky Way
field stars.}.

By contrast, high-resolution spectroscopic measurements for the lower
metallicity range, [Fe/H]$<-3$, are still few in number. Thus, to obtain
stronger constraints on the nature of the r-process, and in particular to
possibly narrow the mass range of supernovae that may be capable of producing
the r-process elements, further investigations for neutron-capture elements in
EMP stars are strongly desired. 

Here we report elemental abundance measurements for two EMP stars discovered
during the course of the Sloan Digital Sky Survey (SDSS; York et al. 2000).
These stars are the first known examples of cool main-sequence stars (subdwarfs)
with [Fe/H]$<-3$, one of which exhibits extremely large excesses of the
r-process elements. 

\section{Observations and Analysis}

\subsection{Sample Selection and Observations}

The two cool subdwarfs investigated in the present work were selected from a
large sample of candidate EMP stars observed during the Sloan Extension for
Galactic Exploration and Understanding (SEGUE; Yanny et al. 2009), based on
atmospheric parameters obtained by the SEGUE Stellar Parameter Pipeline
\citep[SSPP;][]{lee08}. These stars were listed as EMP stars with strong CH
bands and $\log g \sim 3.5$, and were expected to be examples of carbon-enhanced
metal-poor stars.

High-resolution follow-up spectroscopy was conducted with the Subaru Telescope
High Dispersion Spectrograph \citep[HDS;][]{noguchi02} in 2008 August and
October. The ``snapshot'' high-resolution (R=30,000), moderate S/N ($\sim 30$)
observations revealed these two stars to have high surface gravities, based on
the weak absorption lines of ionized species (e.g., \ion{Fe}{2}) and the broad
wings of their \ion{Mg}{1} b lines. The strong CH bands are explained by the
high pressure in the subdwarf's atmospheres, rather than due to a carbon excess.
The metallicity of these two stars are significantly lower than previously known
cool subdwarfs \citep[e.g., ][]{yong03a}, except for the dwarf carbon star
G77-61 \citep{plez05}. Moreover, very strong \ion{Ba}{2} absorption lines were
found for {\sdssa}, suggesting large excesses of the heavy neutron-capture
elements.

\begin{figure}
\includegraphics[scale=.50]{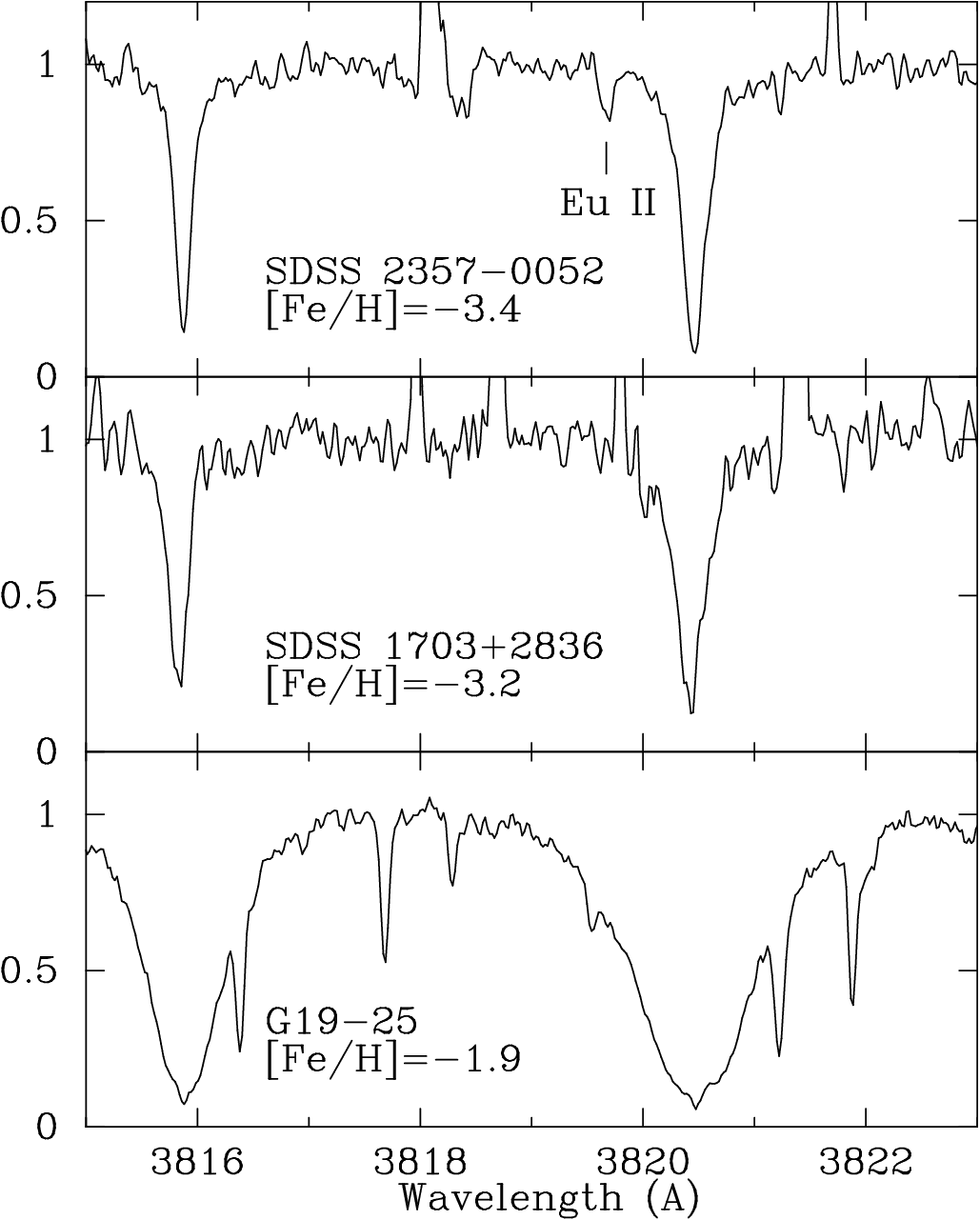}
\caption{Observed spectra around the \ion{Eu}{2} 3819~{\AA} line. The
  stellar name and derived [Fe/H] are given in each
  panel. \label{fig:sp3}}
\end{figure}

In order to determine detailed chemical compositions of these stars, we obtained
higher resolution ($R=60,000$), higher S/N spectra, covering 3500--6800~{\AA},
with Subaru/HDS in 2008-2009. Table~1 lists details of the observations. The
observed spectra around the \ion{Eu}{2} 3819~{\AA} feature are shown in
Figure~\ref{fig:sp3}. For comparison purposes, we also obtained a spectrum with
the same instrumental setup for the cool, very metal-poor ([Fe/H]$\sim -2$)
subdwarf G~19--25, selected from \citet{yong03a}.

\subsection{Data Reduction and Measurements}

Standard data reduction procedures were carried out with the IRAF echelle
package\footnote{IRAF is distributed by the National Optical Astronomy
Observatories, which is operated by the Association of Universities for Research
in Astronomy, Inc. under cooperative agreement with the National Science
Foundation.}. The sky background was removed when the spectra are affected by
the moon, as they were for several of the exposures.

Equivalent widths ($W$'s) for isolated absorption lines were measured
by fitting Gaussian profiles as well as by direct integration of the absorption
features. We found that the values obtained by direct integration are slightly
higher for strong lines, probably due to inadequate fits of the assumed
Gaussian profiles to the line wings.  Hence, we adopted the values
obtained by direct integration when the $\log(W/\lambda)$ value is
larger than $-4.7$, which corresponds to $W=100$~m{\AA} at
$\lambda=5000$~{\AA}, while the results of Gaussian fitting are adopted
for weaker lines. We used the line list compiled by Aoki et al. (2010,
in preparation) for studies of extremely metal-poor stars in the
Galactic halo.

Radial velocities were measured for our program stars using selected Fe lines in
the wavelength region 4050--5200~{\AA}.  Heliocentric radial velocities are
listed in Table~\ref{tab:obs} for individual spectra obtained in each run with
the same setup. The radial velocities of {\sdssa} measured from the
November 2008 and June/July 2009 spectra are about 1~{\kms} different from
measurements obtained during August/October 2008. This difference is
significantly larger than the random errors of the measurements
(Table~\ref{tab:obs}), which are estimated from $\sigma_{\rm v}N^{-1/2}$, where
$\sigma_{\rm v}$ is the standard deviation of the derived values for individual
lines and $N$ is the number of lines used. However, taking the systematic errors
of the measurements into account, which could be as large as 0.5~{\kms}, further
confirmations for the variation of radial velocity are required to derive any
conclusion on the possible binarity of this object.

\subsection{Atmospheric Parameters}

The effective temperatures ({\teff}'s) of our objects are determined
from the $(V-K)_{0}$ colors and the temperature scale of
\citet{alonso96b}, assuming [Fe/H]$=-3.0$ for the SDSS objects.  The
$V_{0}$ magnitude is derived from the SDSS $g_{0}$ and $(g-r)_{0}$,
using the transformations of \citet{zhao06}. The $K$ magnitude is
adopted from the 2MASS catalogue \citep{skrutskie06}, using the
reddening estimates by \citet{schlegel98}. The photometry data and
derived effective temperatures are given in Table~\ref{tab:obs} and
\ref{tab:abund}, respectively. The derived {\teff}'s of the two SDSS
stars are quite similar. In the table, we also give the estimate of
{\teff} obtained from $(g-r)_{0}$ and a temperature scale constructed
based on the ATLAS model
atmospheres\footnote{wwwuser.oat.ts.astro.it/castelli/colors/sloan.html}. These
{\teff}'s are higher by 100--150~K than those from $(V-K)_{0}$ derived
above. We adopt the averages of the two determinations. Application of
the recent effective temperature scale of \citet{casagrande10} yields
temperature estimates about 100~K higher {\teff} than Alonso et al.'s
scale for the same value of $(V-K)_{0}$; this result agrees well with
the values adopted in the present work. The {\teff} of G~19--25 is
adopted from \citet{alonso96a} determined by the Infrared Flux Method,
which was also adopted by \citet{yong03a}.

The surface gravity ($\log g$) for the SDSS stars is estimated from
the isochrones of \citet{kim02} for metal-poor ([Fe/H]$=-3.5$) stars,
adopting ages of 12~Gyr and assuming these stars are on the main
sequence. The surface gravities ({\logg} $=4.8$) are insensitive to
the assumed metallicity and ages for such cool main-sequence
stars. The $\log g$ of G~19--25 is estimated by demanding the Fe
abundances derived from \ion{Fe}{1} and \ion{Fe}{2} agree. The
metallicity adopted in the abundance analysis is [Fe/H]$=-3.3$ and
$-1.9$ for the SDSS stars and G~19--25, respectively, which are
obtained by iterations of the analysis procedure described below.

Elemental abundances are measured by standard analyses for equivalent
widths using ATLAS model atmospheres with no convective overshooting
\citep{castelli03}, with enhancements of the $\alpha$-elements. We
performed analyses for more than 60 \ion{Fe}{1} lines in the SDSS
stars, whose equivalent widths range from 5~m{\AA} to 200~m{\AA}. Our
analyses show that a decreasing trend of derived abundances with
increasing equivalent widths emerges when the van der Waals broadening
based on \citet{unsold55}'s treatment is enhanced, as was done in our
previous studies \citep[e.g. ][]{aoki02}, even though no
micro-turbulence is assumed. Hence, we adopt the \citet{unsold55}'s
broadening with no modification in our analysis. We note that the iron
abundance derived from weak lines ($\log(W/\lambda)<-5$), as well as
the abundances for most of the neutron-capture elements (e.g., Y, La,
Eu), also based on weak lines, are insensitive to the nature of the
pressure broadening.

We confirmed that there remains no significant correlation between the
derived Fe abundances from individual \ion{Fe}{1} lines and their
excitation potential, supporting the estimates of their effective
temperatures from broadband colors.

The absolute magnitude of subdwarfs with $T_{\rm eff}=5000$~K is
$M_{V}$ = 7.4, according to the isochrones of \citet{kim02}.  This
indicates that the stellar mass is 0.52~$M_{\odot}$, and the distances
of {\sdssa} ($V=15.6$) and {\sdssb} ($V=15.3$) are 440~pc and 370~pc,
respectively.  Adopting these distances as well as radial velocities
and proper motions ({\sdssa}: $\mu_{\alpha}, \mu_{\delta} = 46,
-172$~mas yr$^{-1}$; {\sdssb}: $\mu_{\alpha}, \mu_{\delta} = -27,
-129$~mas yr$^{-1}$, with errors of $\sim 2.5$~mas yr$^{-1}$; Munn et
al. 2004), the kinematics of the SDSS stars are calculated using the
procedures of \citet{chiba00}. {\sdssa} is tentatively assigned to the
outer-halo component \citep{carollo07, carollo10}, based on its
retrograde orbit ($V_{\phi}=-83\pm70$~{\kms}), its moderate
eccentricity ($e=0.47$), and its large $Z_{\max}$ value (the maximum
distance of its orbit from the Galactic plane) of 8~kpc, coupled with
its location in a Toomre energy diagram ($V =-306, (U^2+W^2)^{1/2} =
191${\kms}); the kinematics of {\sdssb} are more consistent with the
inner halo.

\subsection{Abundance Measurements}

The derived elemental abundances for our stars are listed in
Table~\ref{tab:abund}. Although only one \ion{Fe}{2} line is clearly
detected in the spectra of the SDSS stars, the derived iron abundance
from the \ion{Fe}{2} line agrees with that from \ion{Fe}{1} lines
within the measurement error. Moreover, the Ti abundances from the two
ionization stages agree well in both objects. These results clearly
exclude the possibility that these stars are cool giants with
{\logg$\sim 2$}, but support the above estimates of the gravities.

The spectrum synthesis technique is adopted to determine the carbon
abundance from the CH molecular band at 4315~{\AA}. Although the
derived abundance ratios are slightly higher than solar, these values
are typical for unevolved, extremely metal-poor stars
\citep[e.g.][]{spite05}, thus these stars are not carbon-enhanced
objects.

Barium abundances are obtained by an analysis including hyperfine
splitting and isotope shifts \citep{mcwilliam98}, assuming the isotope
ratios expected for the r-process-enhanced case (see below). The
final result is derived from the three weaker lines in the red region,
which are less sensitive to hyperfine splitting. Hyperfine splitting is also
included in the analysis of \ion{Mn}{1} lines.

The abundances of Eu, Er, and Yb are determined by the spectrum
synthesis technique, carefully checking on the effects of blending
from other absorption features, for \ion{Eu}{2} 3819, 4129 and
4205~{\AA}, \ion{Er}{2} 3692~{\AA}, and \ion{Yb}{2} 3694~{\AA}. Atomic
line data for Eu, Er, and Yb are adopted from \citet{lawler01},
\citet{lawler08}, and \citet{sneden09}, respectively, including the
effect of hyperfine splitting for Eu and Yb. The upper limit on the Th
abundance is determined by the spectrum synthesis technique for the
\ion{Th}{2} 4019~{\AA} line.

Random errors in the abundance measurements are estimated to be
$\sigma N^{-1/2}$, where $\sigma$ is the standard deviation of derived
abundances from individual lines and $N$ is the number of lines used
in the analysis. When only a few lines are available, the $\sigma$ of
\ion{Fe}{1} is adopted in the estimates.  The
errors due to the uncertainty of the atmospheric parameters
($\delta${\teff}$=150$~K, $\delta${\logg}$=0.3$,
$\delta${\vt}$=0.2$~{\kms}) are also estimated, and added in
quadrature to the random errors.
The metallicity of G~19--25 obtained by our analysis agrees, within
the errors, with that derived by \citet{yong03a}, adopting the same
{\teff}.

\section{Discussion}

The two SDSS stars studied in the present work turned out to be the
lowest metallicity cool subdwarfs yet known, with
[Fe/H]$<-3$. Metal-poor subdwarfs have been extensively studied by
\citet{yong03a}, yet their most metal-poor objects have [Fe/H]$\sim
-2.5$. The much larger, and deeper SDSS/SEGUE surveys are clearly
potentially interesting sources for the discovery of additional
examples of extremely metal-poor, cool subdwarfs.

These two stars exhibit quite similar abundance ratios of their
lighter elements ($Z\leq 28$). The over-abundances of the
$\alpha$-elements and under-abundances of Mn are similar to those
found for other EMP stars. Other iron-peak elements trace the Fe
abundances, while EMP giants usually show under-abundances of Cr as
well as over-abundances of Co. However, discrepancies of Cr abundance
ratios between EMP giants and main-sequence turn-off stars are known
to occur \citep[e.g. ][]{lai08}, and our results are similar to those
of turn-off stars.  Lithium is not detected for either star, and the
upper limits on its abundance are significantly lower than the Spite
plateau value, as expected for cool stars with substantial convective
envelopes.

\begin{figure}
\includegraphics{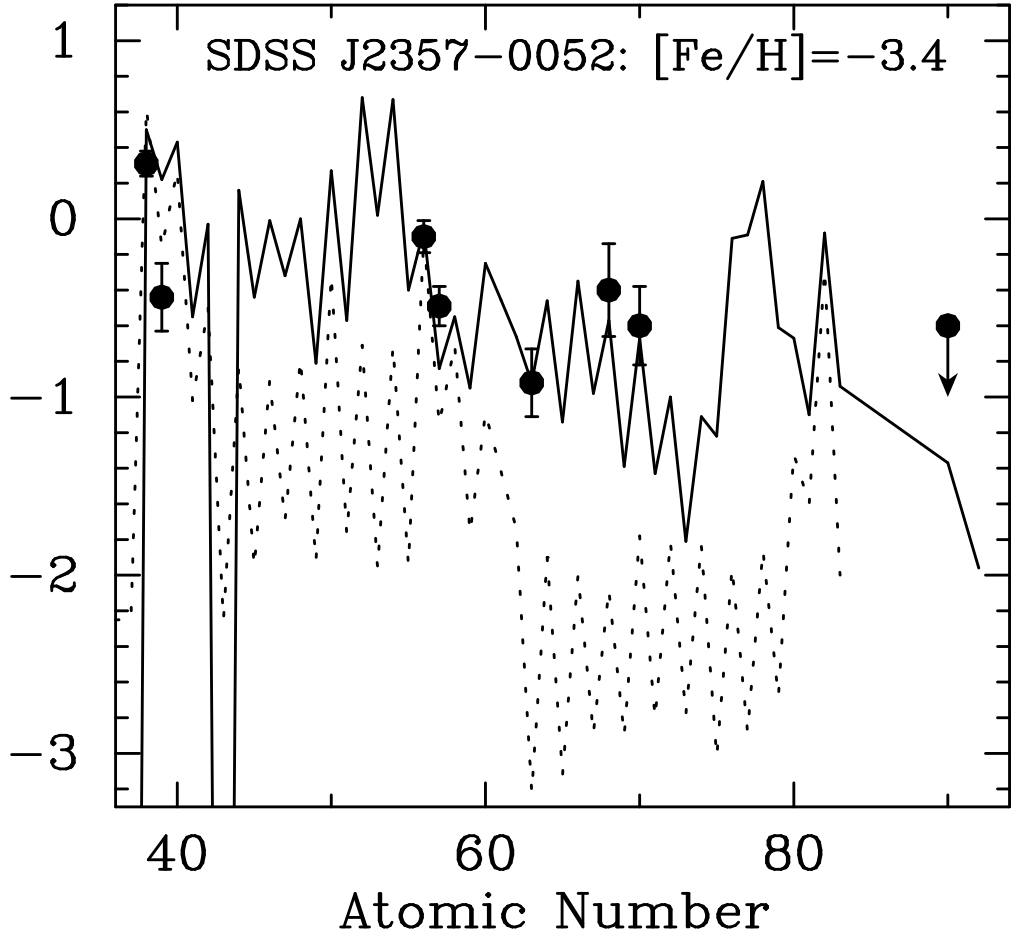}
\caption{The abundances ($\log \epsilon$ values) of neutron-capture
  elements in {\sdssa} (dots). The solid and dotted lines indicate
  the abundance patterns of the r- and s-process components in
  Solar System material \citep{arlandini99}, normalized at Ba ($Z=56$). 
\label{fig:pattern}}
\end{figure}

A surprising result is that one of the first two subdwarfs with
[Fe/H]$<-3$ studied by high-resolution spectroscopy exhibits large
excesses of neutron-capture elements. Figure~\ref{fig:pattern} shows
the abundances ($\log \epsilon$ values) of neutron-capture elements in
{\sdssa}. Comparisons with the abundance patterns of the r- and
s-process components in solar-system material indicate that the origin
of neutron-capture elements is clearly associated with the r-process.
This result immediately confirms that large excesses of r-process
elements are not a phenomena found only for red giants \citep[see
][]{sneden08}.

\begin{figure}
\includegraphics{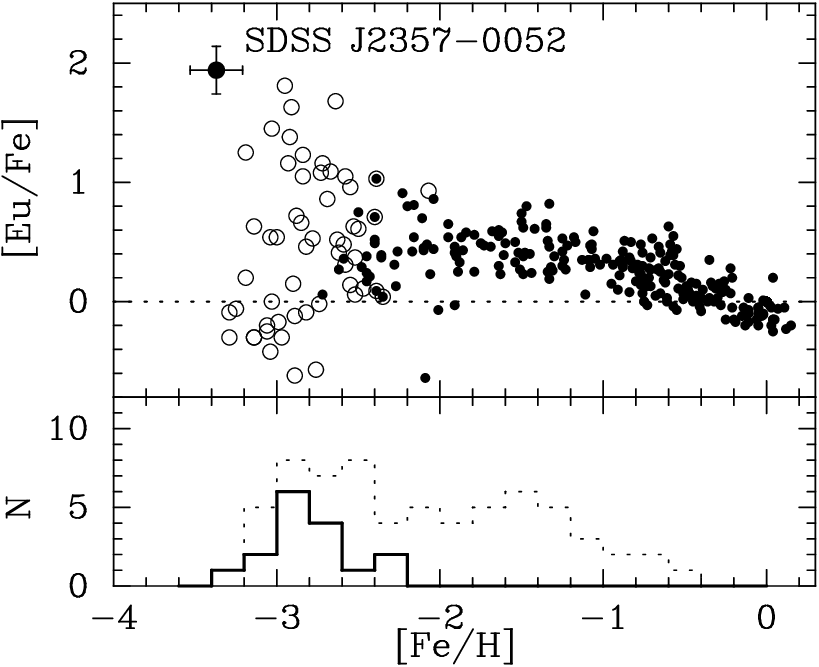}
\caption{The Eu abundance ratios of Galactic halo and disk stars
  (upper panel). The large filled circle indicates our result for
  {\sdssa}, while open circles are for very metal-poor stars
  taken from literature \citep{hill02, sneden03, honda04, ishimaru04,
    christlieb04, aoki05, barklem05, preston06, francois07, frebel07,
    hayek09, roederer10}.  Results for less metal-poor stars shown by
  the small filled circles are taken from the SAGA database
  \citep{suda08}. The lower panel shows the distribution of r-II stars
  ([Eu/Fe]$>+1.0$, solid line) and r-II + r-I ([Eu/Fe]$>+0.5$) stars (dotted
  line), following the nomenclature of Beers \& Christlieb (2005).
 \label{fig:eufe}}
\end{figure}

The [Eu/Fe] abundance ratio of {\sdssa} ([Eu/Fe]$=+1.9$) is the
highest among the stars that exhibit excesses of r-process elements at
low metallicity (Fig.  \ref{fig:eufe}). This is basically because of
the low Fe abundance of this object ([Fe/H]$=-3.4$). Indeed, the Eu
abundance of this object ([Eu/H] $=-1.4$) is similar to those
previously found for other r-II stars, e.g., CS~31082-001:
[Eu/H]$=-1.28$ \citep{hill02}; HE~1523--0901 : [Eu/H]$=-1.14$
\citep{frebel07}.

The large excesses of heavy neutron-capture elements in r-II stars are
usually interpreted as a result of the pollution of interstellar
matter by a supernova (or some other event) that yields amounts of
r-process elements and subsequent low-mass star formation from the
enriched material. The fact that only a small fraction of stars show
such excesses (roughly 5\% of stars with [Fe/H] $< -2$ according to
Barklem et al. 2005) indicates that the event(s) associated with the
production of r-process elements were rare, and did not produce
significant amounts of Fe.

Since r-II stars have previously only been found in the metallicity
range [Fe/H]$\gtrsim-3$ (Fig. 3), the respective sites of the
r-process have been supposed to be supernovae of less massive
progenitors (e.g. 8--10M$_{\odot}$; Wanajo \& Ishimaru 2006). Our
discovery of {\sdssa} indicates that the metallicity range in which
highly r-process-enhanced stars appear extends to as low as
[Fe/H]$=-3.4$, and thus opens the possibility that a larger range of
progenitor masses may have been involved.

The apparent upper bound of [Eu/H] in r-II stars might indicate the
existence of a limit of the enrichment by a supernova yielding
r-process elements. The mass ratio of Eu/H corresponding to
[Eu/H]$\sim -1.5$ is $M_{\rm Eu}/M_{\rm H}\sim 10^{-11}$. If the Eu
mass produced by a single supernova is assumed to be
$10^{-7}$M$_{\odot}$ \citep{wanajo06}, the amount of interstellar
matter polluted by that could only be on the order of
$10^{4}$M$_{\odot}$. The Fe production by such stars is estimated to
be $\sim 2\times10^{-3}$M$_{\odot}$ \citep{wanajo09}. If this is mixed
into a metal-free cloud of $10^{4}$M$_{\odot}$, the resulting
metallicity is [Fe/H]$\sim -4$. If the cloud mass is assumed to be
smaller, taking account of the low explosion energy of a supernova by
a less massive progenitor, the Fe abundance of {\sdssa}, as well as
those of neutron-capture elements, could be explained by the
contribution of a single supernova.

Another scenario to explain the large excesses of r-process elements
in {\sdssa} is that the object belongs to a binary system in which the
massive companion has exploded, yielding the r-process elements. A
possible binary system with a highly elliptical orbit is suggested for
the r-process-enhanced star HE~2327--5642 ([Fe/H] $= -2.78$; [Eu/Fe] =
+0.98) by \citet{mashonkina10}, based on a single radial velocity
measurement showing significant departure ($\sim 20$ {\kms}) from 16
others (with errors on the order of 0.2-0.5 {\kms}). Since this
observation (if not spurious) may have some relationship with the
phenomena of large r-process excess, further investigations of radial
velocity changes for r-II stars (including {\sdssa}) are strongly
desired.

It may be of interest that the first extremely metal-poor r-II
subdwarf appears to be associated with the outer halo.  Given the
possible distinct astrophysical origins between the inner- and
outer-halo populations (e.g., masses of their progenitor sub-haloes,
and/or star formation histories; Carollo et al. 2007, 2010), it would
clearly be desirable to establish membership assignments for the known r-II
stars. This can be realized with sufficient accuracy if
additional r-II subdwarfs are found. This could be accomplished
by targeting candidate very and extremely metal-poor stars identified by 
SDSS/SEGUE with high proper motions (likely subdwarfs) and low {\teff}.

\acknowledgments




{\it Facilities:} \facility{SDSS}, \facility{Subaru(HDS)}.

\acknowledgments

Funding for the SDSS and SDSS-II has been provided by the Alfred P. Sloan
Foundation, the Participating Institutions, the National Science Foundation, the
U.S. Department of Energy, the National Aeronautics and Space Administration,
the Japanese Monbukagakusho, the Max Planck Society, and the Higher Education
Funding Council for England. The SDSS Web Site is http://www.sdss.org/.

The SDSS is managed by the Astrophysical Research Consortium for the
Participating Institutions. The Participating Institutions are the American
Museum of Natural History, Astrophysical Institute Potsdam, University of Basel,
University of Cambridge, Case Western Reserve University, University of Chicago,
Drexel University, Fermilab, the Institute for Advanced Study, the Japan
Participation Group, Johns Hopkins University, the Joint Institute for Nuclear
Astrophysics, the Kavli Institute for Particle Astrophysics and Cosmology, the
Korean Scientist Group, the Chinese Academy of Sciences (LAMOST), Los Alamos
National Laboratory, the Max-Planck-Institute for Astronomy (MPIA), the
Max-Planck-Institute for Astrophysics (MPA), New Mexico State University, Ohio
State University, University of Pittsburgh, University of Portsmouth, Princeton
University, the United States Naval Observatory, and the University of
Washington.

W.~A. would like to acknowledge useful discussions on the r-process
with S. Wanajo and Y. Ishimaru.  W.~A. is supported by a Grant-in-Aid
for Science Research from JSPS (grant 18104003). T.C.B. acknowledges
partial funding of this work from grants PHY 02-16783 and PHY
08-22648: Physics Frontier Center/Joint Institute for Nuclear
Astrophysics (JINA), awarded by the U.S. National Science
Foundation. D.C. acknowledges funding from RSAA ANU to pursue her
research.

\clearpage






\clearpage

\begin{deluxetable}{llccccc}
\tabletypesize{\scriptsize}
\tablecaption{OBJECTS AND OBSERVING LOG \label{tab:obs}}
\tablewidth{0pt}
\tablehead{
\colhead{Star\tablenotemark{a}} & \colhead{Obs. date} & \colhead{Setup\tablenotemark{b}} & \colhead{Exposure}  & \colhead{Count} &
\colhead{HJD} & \colhead{\vh} \\
\colhead{} & (UT) & & (min.) & (5000~{\AA})& & ({\kms})
}
\startdata
{\sdssa} & Aug 21, 2008 & Yd &15 & 430 & 2454700 & $-9.1\pm0.3$ \\
$V_{0}=15.612$ & Oct 4, 2008 & Yd & 15 & 480 & 2454744& $-9.3\pm0.1$\\
$(V-K)_{0}=2.059$ & Nov. 16, 2008 & Bc & 160 & 2890 & 2454786 & $-10.0\pm0.1$ \\
$(g-r)_{0}=0.612$ & June 30/July 2, 2009 & Bc & 180 & 2610 & 2455012/14 & $-10.2\pm0.1$ \\
$E(B-V)=0.030$ & Sep 11, 2009 & Yd & 80 & 1620 & 2455086 & $-9.0\pm0.1$ \\
         & Sep 12, 2009 & Bc & 120 & 1690 & 2455087 & $-9.2\pm0.1$ \\
\hline
{\sdssb}  & July 6,2008 & Yd & 10 & 390 & 2454654 & $-178.2\pm0.2$\\
$V_{0}=15.342$ & July 2, 2009 & Bc & 120 & 1650 & 2455015 & $-178.0\pm0.1$\\
$(V-K)_{0}=2.098$ & Sep 11, 2009 & Yd & 120 & 1080 & 2455086 & $-178.1\pm0.1$\\
$(g-r)_{0}=0.593$         & Sep 12, 2009 & Bc & 120 & 1970 & 2455087 & $-177.7\pm0.1$\\
$E(B-V)=0.065$ & \nodata & \nodata  & \nodata&\nodata&\nodata&\nodata\\
\hline
G~19--25 &  Sep 10, 2009 & Yd & 10 & 7700 & 2455086 & $-32.4\pm0.2$ \\
$V_{0}=11.64$ & Sep 11, 2009 & Bc & 10 & 8000 & 2455087 & $-32.5\pm0.1$ \\
$(V-K)_{0}=2.023$ & \nodata & \nodata & \nodata & \nodata & \nodata & \nodata \\
$E(B-V)=0.000$ & \nodata & \nodata & \nodata & \nodata & \nodata & \nodata \\
\enddata
\tablenotetext{a}{The full names are SDSS J235718.91$-$005247.8 and SDSS J170339.60$+$283649.9.
The PLATE-MJD-FIBER names in the SDSS/SEGUE databases are 1489-52991-251 and
2808-54524-510, respectively.}
\tablenotetext{b}{The HDS standard setups Yd and Bc cover 4100--6800~{\AA} and 3550--5250~{\AA}, respectively}
\end{deluxetable}

\begin{deluxetable}{lccccccccccccccc}
\tabletypesize{\scriptsize}
\tablecaption{STELLAR PARAMETERS AND CHEMICAL ABUNDANCE RESULTS \label{tab:abund}}
\tablewidth{0pt}
\tablehead{
\colhead{} & \multicolumn{4}{c}{\sdssa} && \multicolumn{4}{c}{\sdssb} &&  \multicolumn{4}{c}{G~19--25}}
\startdata
{\teff}$(V-K)$(K) & \multicolumn{4}{c}{4960} && \multicolumn{4}{c}{4908} & & \multicolumn{4}{c}{4990} \\
{\teff}$(g-r)$(K) & \multicolumn{4}{c}{5050} && \multicolumn{4}{c}{5100} & & \multicolumn{4}{c}{\nodata} \\
{\teff}(K)        & \multicolumn{4}{c}{5000} && \multicolumn{4}{c}{5000} & & \multicolumn{4}{c}{4900} \\
{\logg}           & \multicolumn{4}{c}{4.8}  && \multicolumn{4}{c}{4.8}  & & \multicolumn{4}{c}{4.6}  \\
{\vt}({\kms})     & \multicolumn{4}{c}{0.0}  && \multicolumn{4}{c}{0.0}  & & \multicolumn{4}{c}{0.2} \\
\hline
 element  & $\log \epsilon$ & [X/Fe] & N & error &&$\log \epsilon$ & [X/Fe] & N & error &&$\log \epsilon$ & [X/Fe] & N & error \\
\hline
Fe (Fe I) & 4.14 & $-3.36$ & 78 & 0.16 &&4.27 & $-3.23$ & 63 & 0.16 && 5.63 & $-1.87$ & 149 & 0.17  \\
Fe (Fe II)& 4.00 & $-0.15$ & 1  & 0.25 &&4.21 & $-0.06$ & 1  & 0.24 && 5.66 & 0.03 & 11 & 0.26  \\
Li (Li I) &$<1.0$&\nodata&\nodata&\nodata&&$<1.0$&\nodata&\nodata&\nodata&&$<0.2$ &\nodata &\nodata &\nodata \\
C (CH)    & 5.5  & 0.43    & \nodata & 0.11 && 5.4 & 0.20 &\nodata& 0.11 && 6.3 & $-0.26$ &\nodata  & 0.11  \\
Na (Na I) & \nodata &\nodata&\nodata&\nodata&& 2.85 & $-0.17$ & 2 &  && 3.83 & $-0.54$ & 1 & 0.17  \\
Mg (Mg I) & 4.43 & 0.19    & 3  & 0.12 &&4.58 & 0.21 & 2 & 0.12 && 5.97 & 0.23 & 4 & 0.05  \\
Si (Si I) & 4.90 & 0.75    & 1  & 0.19 &&\nodata&\nodata&\nodata&\nodata&& 6.12 & 0.48 & 1 & 0.21  \\
Ca (Ca I) &\nodata&\nodata&\nodata&\nodata&&3.29 & 0.18 & 2 & 0.06 && 4.65 & 0.18 & 18 & 0.03  \\
Ca (Ca II) & 3.37 & 0.38   &  1 & 0.22 &&3.00 & $-0.11$ & 1 & 0.21 &&\nodata&\nodata&\nodata&\nodata \\
Sc (Sc II) &$-0.07$& 0.14 & 1 & 0.18 &&\nodata&\nodata&\nodata&\nodata&& 1.35 & 0.07 & 2 & 0.18  \\
Ti (Ti I) & 1.79 & 0.19 & 3 & 0.06 &&1.90 & 0.18 & 2 & 0.08 && 3.41 & 0.33 & 18 & 0.08  \\
Ti (Ti II) & 1.73 & 0.14 & 6 & 0.13 &&1.84 & 0.12 & 4 & 0.15 && 3.55 & 0.47 & 15 & 0.18  \\
Cr (Cr I) & 2.14 & $-0.14$ & 5 & 0.05 &&2.43 & 0.02 & 6 & 0.04 && 3.80 & 0.02 & 8 & 0.05  \\
Mn (Mn I) & 1.59 & $-0.48$ & 2 & 0.12 &&1.87 & $-0.33$ & 3 & 0.12 && 3.14 & $-0.42$ & 6 & 0.20  \\
Co (Co I) & 1.72 & 0.09 & 4 & 0.06 && 1.80 & 0.04 & 1 & 0.10 && 3.08 & $-0.04$ & 7 & 0.05  \\
Ni (Co I) & 2.96 & 0.09 & 4 & 0.04 && 3.10 & 0.11 & 5 & 0.06 && 4.24 & $-0.12$ & 8 & 0.12  \\
Sr (Sr II) & 0.29 & 0.78 & 2 & 0.07 &&$-1.18$& $-0.82$ & 1 & 0.09 && 0.97 & $-0.04$ & 2 & 0.12  \\
Y (Y II)  & $-0.44$& 0.71 & 1 & 0.19 &&\nodata&\nodata&\nodata&\nodata && 0.40 & 0.06 & 4 & 0.22  \\
Zr (Zr II) &\nodata&\nodata&\nodata&\nodata&&\nodata &\nodata &\nodata &\nodata &&\nodata &\nodata &\nodata &\nodata  \\
Ba (Ba II) &$-0.10$& 1.12 & 4 & 0.09 &&\nodata &\nodata &\nodata& \nodata && 0.14 & $-0.18$ & 3 & 0.12  \\
La (La II) &$-0.49$& 1.08 & 2 & 0.11 &&\nodata &\nodata &\nodata&\nodata  &&$-0.36$ & 0.40 & 1 & 0.26  \\
Eu (Eu II) &$-0.92$& 1.92 & 3 & 0.19 &&\nodata &\nodata &\nodata&\nodata  &&\nodata &\nodata &\nodata &\nodata  \\
Er (Er II) &$-0.40$& 2.04 & 1 & 0.26 &&\nodata &\nodata &\nodata&\nodata  &&\nodata &\nodata &\nodata&\nodata  \\
Yb (Yb II) &$-0.60$& 1.92 & 1 & 0.22 &&\nodata &\nodata &\nodata&\nodata  &&\nodata &\nodata &\nodata &\nodata  \\
Th (Th II) &$<-0.60$&$<2.74$ &\nodata&\nodata&&\nodata &\nodata & \nodata&\nodata &&\nodata &\nodata &\nodata &\nodata  \\
\hline
\enddata
\end{deluxetable}


\end{document}